\documentclass{llncs}
\usepackage{amsmath,graphicx}
\usepackage{multirow}

\begin{document}
\title{Automated Detection and Type Classification of Central Venous Catheters in Chest X-rays}
\titlerunning{Automated Catheter Detection and Classification}

\author{Vaishnavi Subramanian\inst{1,2}, Hongzhi Wang\inst{1}\thanks{Corresponding author.}   \and Joy T. Wu\inst{1} \and \\Ken C. L. Wong\inst{1} \and Arjun Sharma\inst{1} \and Tanveer Syeda-Mahmood\inst{1}}

\institute{IBM Research, Almaden Research Center, San Jose, CA, USA\\ \and University of Illinois Urbana-Champaign, Urbana, IL, USA \\
\email{hongzhiw@us.ibm.com}
}

%
\maketitle

\begin{abstract}
\label{sec:abstract}
Central venous catheters (CVCs) are commonly used in critical  care  settings for monitoring body functions and administering medications.  They are often described in radiology reports by referring to their presence, identity and placement. In this paper, we address the problem of automatic detection of their presence and identity through automated segmentation using deep learning networks and classification based on their intersection with previously learned shape priors from clinician annotations of CVCs. The results not only outperform existing methods of catheter detection achieving 85.2\% accuracy at 91.6\% precision, but also enable high precision (95.2\%) classification of catheter types on a large dataset of over 10,000 chest X-rays, presenting a robust and practical solution to this problem. 
\end{abstract}

\section{Introduction}
\label{sec:intro}
Central venous catheters (CVCs) are commonly used in critical care settings and surgeries to monitor a patient's heart function and deliver medications close to the heart. These are inserted centrally or peripherally through the jugular, subclavian or brachial veins and advanced towards the heart through the venous system, most often blindly. Portable anterior-posterior (AP) chest X-Rays (CXRs) obtained after the CVC placements are used to rule out malpositioning and complications. The interpretation of the CXRs is currently done manually after loading these into the hospital's electronic systems. 

\begin{figure}[t]
    \centering
    \includegraphics[width=\textwidth]{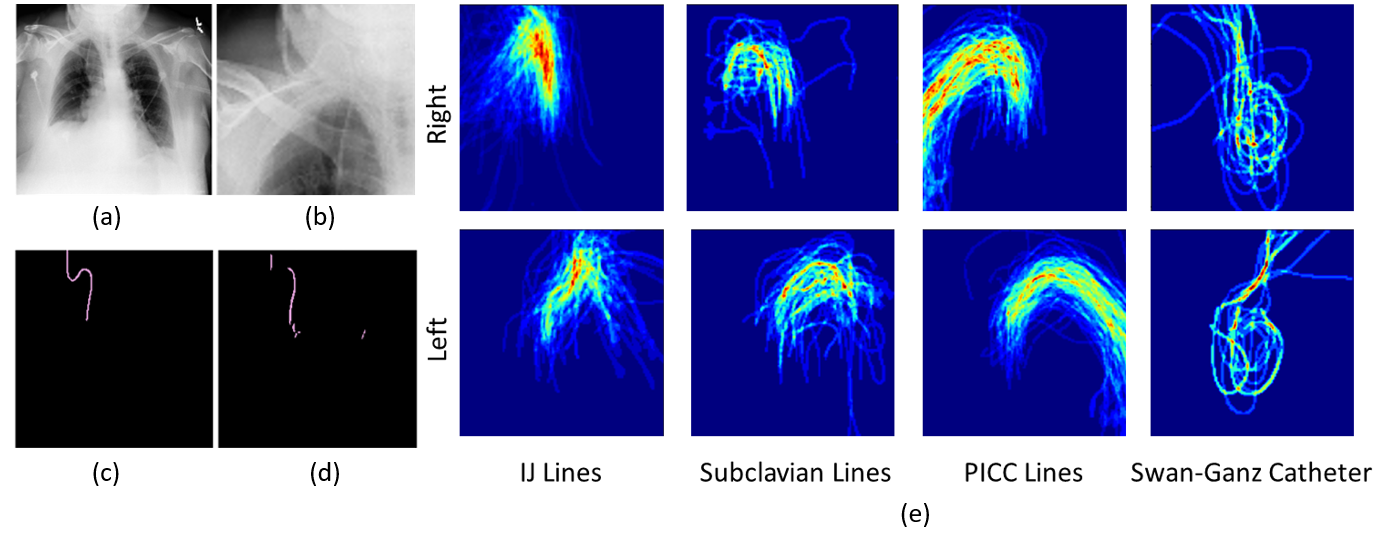}
    \caption{Example of central venous catheters (CVCs) and collection of annotations:  (a)~Original CXR where the CVC is barely visible, (b) enlarged region of interest focusing on the CVC, (c) manual CVC annotation, (d) segmentation produced by U-Net (e) The pixel-wise overlay of manual annotations for four common types of CVCs, which are used to construct spatial priors.}
    \label{fig:example}
\end{figure}

With the advancement of deep learning approaches to anatomical findings in chest X-rays, it is conceivable that radiology reports may be produced automatically in the future, which would considerably expedite the clinical workflow. However, any such report, particularly for in-hospital settings, will need to mention the presence of the CVC, its type such as internal jugular (IJ) or peripherally inserted central catheter (PICC), and any problems with their positioning and insertions. Different types of CVCs also have slightly different optimal tip locations. Automated detection and recognition of CVCs through direct whole image based recognition approaches is unlikely to yield good results as it is difficult to learn discriminative features from these thin tubular structures that occupy less than 1\% of the footprint in the overall image, as shown in Fig.~\ref{fig:example}(a-d). Hence most methods have focused on local extraction of these structures. 

There is vast literature using conventional medical image processing on the detection of catheters both in angiography imaging and chest X-rays. Recent work, however, has applied deep learning for the detection of the presence of the catheters and their tips. In~\cite{lee2017deep}, the detection of tip location for PICC lines was attempted. In CXR fluoroscopy images, the sequence information was used to aid automatic segmentation of catheters~\cite{ambrosini2017fully} based on the U-Net~\cite{ronneberger2015u}. An approach recently proposed~\cite{yi2019automatic} for segmentation employs scale-recurrent neural networks on synthetic catheters in pediatric data. Existing approaches result in partial detection using the deep learning stages with post-processing steps to complete the contours, have been tested on smaller datasets or on synthetic datasets.
Their robustness to variations on real, large datasets still needs to be determined. Further, while existing methods aim for either segmentation and detection of a specific CVC type or its placement, the general task of identifying the type of CVC shown in adult chest X-rays has not been attempted, to our knowledge. 

In this paper, we simultaneously address the detection and classification of the CVC type on a large public CXR dataset. Specifically, we can detect and distinguish between four common types of CVCs, namely,  peripherally inserted central catheters (PICC), internal jugular (IJ), subclavian and Swan-Ganz catheters. Our main idea is to augment the detection of catheters with the introduction of shape priors generated from clinician segmentation of catheters to focus on relevant regions for classification. 

Recognizing the identity of the catheters can be a challenge since different CVCs have different contours depending on the origin of insertion and how far they go into the body. However, after annotation of these catheters on hundreds of AP chest X-rays by our clinicians, we discovered that the shape spanned by the various CVC insertion approaches still have a surprisingly distinctive signature and proximity patterns to known anatomical structures as shown in Fig.~\ref{fig:example}(e). Since the type of CVCs can be easily recognized visually by the insertion approaches, in this paper, we exploit this information to both confirm the presence of CVCs as well as recognize their identity.  
Specifically, we first detect the presence of CVCs using prior annotations of these structures through a deep learning segmentation network based on U-Net~\cite{ronneberger2015u,wong20183d} to approximately identify potential fragments of CVCs if present. We then develop features exploiting the segmentation within regional priors and their relation to key anatomical features. The features are then classified into the respective labels of CVCs using random forests. By utilizing the spatial priors, we demonstrate that the complete delineation of the CVC is not required for recognizing either its presence or type. We demonstrate that our features are superior to features extracted by deep learning methods such as VGG16 and DenseNet~\cite{simonyan2014very,huang2017densely} in a comparative study for this problem.
Therefore, our approach is a hybrid combination of automatic feature learning for initial candidate CVC regions with high precision custom features for recognizing the identity of the CVCs based on prior knowledge, in the form of shape priors, within a conventional machine learning classifier. 

\section{Method}
\label{sec:method}
\begin{figure}[t]
    \centering%
    \includegraphics[width=\textwidth]{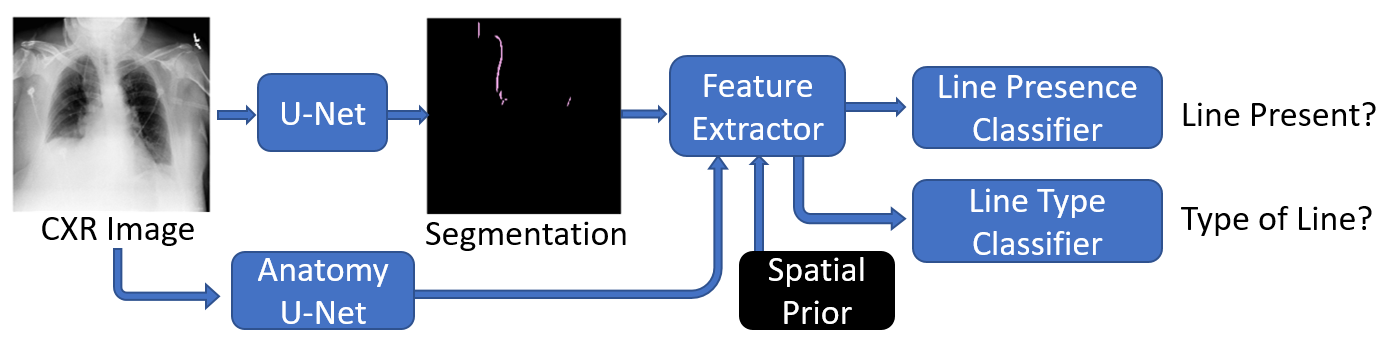} %
    \caption{Our workflow for CXR comprising U-Net based CVC segmentation, feature extraction using spatial prior and anatomies, and random forest based classifiers.}
    \label{fig:workflow}%
\end{figure}
We now describe our hybrid approach to detecting the presence of CVCs and the classification of their types in AP chest X-ray images, shown in Fig.~\ref{fig:workflow}.

\subsection{Segmentation of CVC using modified U-Net}
\label{method:u_net}
To identify regions of interest, we adapted the commonly used U-Net~\cite{ronneberger2015u} for CVC segmentation. While the original loss function works well for large structures, because CVCs are small structures, we used the exponential logarithmic loss function proposed recently~\cite{wong20183d} to address the highly imbalanced label sizes. When the CXR has no CVC of interest, the segmentation output is a blank image, representing no interesting region.

\subsection{Feature Extraction}
\label{sec:feature_ip}
Given the segmentation of possible CVCs in the CXR from the U-Net, we design image processing features describing the segmentation, its relation to prior knowledge of CVC contours, and its relation to key anatomical features of the chest as below. These describe the overall properties of the potential CVCs, even if the initial detected region is imperfect.

\textbf{Spatial prior for each class}: To obtain shape priors, our clinicians annotated the contours of CVCs in hundreds of training images. The CVCs are traced from their anatomical origin of insertion to their tip to give contours that are reflective for different types of CVCs. We averaged the manual annotations per-pixel for each class~(Fig.~\ref{fig:example}(e)) and blurred them spatially to obtain signature spatial priors for each of the CVC classes: left/right PICC, IJ, subclavian and Swan-Ganz catheters as shown in Fig.~\ref{fig:example}(e).

We then performed a pixel-wise multiplication of the segmentation output with the prior for each class and characterized this overlap using an $n$-bin intensity histogram, and histogram of oriented gradients (HoG) features for each class. This informs us how well a CVC segmentation aligns with priors of particular CVC classes.

\textbf{Relation to anatomical features}: 
We extracted the segmentation of chest anatomical structures including the clavicles, lungs, heart and mediastinum using an independent U-Net again trained on clinician annotations. We obtained the Euclidean distance distributions of the segmentation relative to the center of these different chest anatomies. 
These provide contextual information and distinguish confusing classes such as PICC and subclavian lines. 

\textbf{Size and shape properties}: We also characterized the shape and size properties of the positive regions of segmentation: the area, length and width characterize the overall presence/absence of CVCs, the histogram of oriented gradients (HoG) describe the shape contours which are crucial for type classification since different CVC types have different shape signatures, as shown in Fig.~\ref{fig:example}(e).

\subsection{Classification using Random Forests}
The extracted features are used for classification on the presence of CVCs, and identifying the type of CVCs. 
We employ a random forest (RF) for each task. The first RF yields a binary presence/absence label, and the second provides a multi-label output, with 4 indicators, one for each type of CVC: PICC, IJ, subclavian and Swan-Ganz. \\

We chose the current architecture after experimenting with different end-to-end deep learning architectures. Feeding spatial priors and anatomical segmentation as additional channels into a VGG-like network failed to learn distinguishing relations. For our specific problem characterized by small area footprint, long and indistinguishable tubular structures with uneven sample sizes, random forests were found to have better generalization. 

\section{Experiments and Results}
\subsection{Data}

We worked with the NIH dataset comprising of 112,000 CXRs~\cite{NIHcxr_wang2017chestx}. A subset of these were labelled and annotated for different tasks:

\textbf{CVC Segmentation - Pixel level annotation:}
A random sample of 1500 AP CXRs was selected from the NIH dataset. The CXRs with CVCs were annotated at the pixel level, to provide annotations for
359 IJ lines, 78 subclavian lines, 277 PICC lines, and 32 Swan-Ganz catheters, yielding a total of 608 annotated images of size 512 x 512. The remaining images have no lines. An overlay of the different annotations, based on CVC types, is shown in Fig.~\ref{fig:example}(e).

\textbf{CVC Presence - Whole image level annotation:}
A subset of 3000 images was chosen from the NIH dataset, which a radiologist labeled globally for presence of external medical devices. This resulted in 2381 CXRs with some device present, and 619 CXRs with an absence of any device. Since devices are usually associated with catheters, this label extends to CVC presence. 

\textbf{CVC Type - Whole image level annotation:}
A subset of around 16,000 CXRs was sampled from the NIH dataset and a group of radiologists reported on these using a semi-structured template, where one section was dedicated for device reporting. 
A recently proposed NLP sentence clustering algorithm~\cite{bechmarkISBI2019} was applied to the device section and a sentence from each resultant cluster was validated manually by two clinicians to derive the global labels for the presence of different CVCs. 

This resulted in 10,746 CXRs with at least one type of externally inserted catheter, with 4249 PICC lines, 1651 IJ lines, 201 subclavian lines, 192 Swan-Ganz catheters, and 4453 CXRs with other catheters including airway and drainage tubes.
The related dataset has been released as part of the MICCAI 2019 Multimodal Learning for Clinical Decision Support (ML-CDS) Challenge.

\subsection{U-Net Segmentation}
\label{sec:unet_results}

\begin{figure}[t]
\centering
    \includegraphics[width=\textwidth]{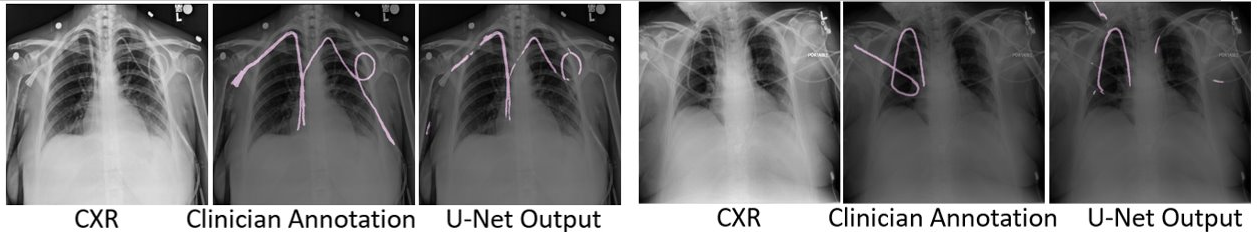}
    \caption{Some example CXR images, overlay of clinician annotations of CVCs, and overlay of U-Net segmentations on the original CXR.}
\label{fig:unet}
\end{figure}

We trained a U-Net for identifying the lines as discussed in Sec.~\ref{method:u_net}, treating the all clinician annotated CVCs as belonging to a positive class without specific distinction on CVC type. We split the pixel-level annotated images into 80\% training and 20\% validation.
We trained the U-Net until convergence using the Adam optimizer with learning rate of $5e^{-5}$ and the exponential loss with the best-performing weights ($w_{\text{Dice}} = 0.8 $ and $w_{\text{Cross}} = 0.2 $)  from~\cite{wong20183d}.

Some segmentation results are shown in Fig.~\ref{fig:unet}.
We observe that sections of the CVCs are missed in some cases, and there are false positives in some confusing areas of the CXRs. 

We evaluate the quality of our segmentation by computing the extent of overlap between the ground truth annotations and the U-Net segmentation output on the held-out set. Since CVCs are thin structures, for reliable overlap estimation between automatic and manual CVC segmentation, we enlarged the binary manual segmentations via a dilation operation. With the radius of 2-pixel dilation, 75\% cases have $>$50\% overlap and 84\% cases have $>$40\% overlap and a 5-pixel dilation radius resulted in 80\% and 90\% cases with greater than 50\% and 40\% overlap, respectively.

\setlength{\tabcolsep}{3.5pt}
\begin{table}[t]
\centering
\caption{Results for CVC detection (P: Precision, R: Recall, Acc: Acccuracy, AUC: Area under ROC, DN: DenseNet, VGG:VGG16, SP: spatial prior). Best values for each column are in bold.}
\resizebox{\textwidth}{!}{
\begin{tabular}{l|cccc|l|cccc}
\hline
Method & P & R & Acc & AUC & Method & P & R & Acc & AUC\\ \hline
1. DN & {20.0} & {20.0} & {20.0} & {50.0} & 8. Seg-VGG-RF & {80.2} & {95.0} & {78.2} & {53.6} \\
2. VGG & {20.0} & {20.0} & {20.0} & {50.0} & 9. + CXR & {79.2} & {97.2} & {79.8} & {49.8} \\
3. DN+VGG & {81.6} & {95.0} & {75.3} & {63.0} & 10. Mask-DN-RF & {84.0} & {92.4} & {79.4} & {62.0} \\
4. DN-RF & 79.4 & \textbf{98.8} & 78.8 & 50.2 & 11. Mask-VGG-RF & 84.2 & 91.6 & 79.6 & 62.8 \\
5. VGG-RF & 79.4 & 98.0 & 78.4 & 50.0 & 12. Seg-SP-RF (Ours) & 89.6 & {89.6} & {83.8} & 75.0 \\
6. Seg-DN-RF & 80.0 & 95.8 & 77.8 & 58.6 & 13. + HoG (Ours) & 91.4 & 89.4 & \textbf{85.2} & 78.8 \\
7. + CXR & 79.6 & 97.2 & 78.4 & 51.4 & 14. + Anatomy (Ours) & \textbf{91.6} & {89.6} & \textbf{85.2} & \textbf{79.4} \\ \hline
\end{tabular}
}
\label{table:presence}
\end{table}

\subsection{CVC Presence Identification}
We next focus on the task of identifying CVC presence in CXRs. We performed a 5-fold cross-validation using a 60-20-20 split for training-validation-testing using the 3000 CXRs labeled for device presence. The classifier outputs a binary label indicating the presence of at least one CVC in the CXR (label: 1), or the absence of any CVCs (label: 0). Results for this task are presented in Table~\ref{table:presence}. The parameters (number of trees, and depth of trees) for all random forests were chosen with hyper-parameter tuning for validation performance, for each fold.

To set the baseline, we trained state-of-the-art VGG16~\cite{simonyan2014very} and  DenseNet~\cite{huang2017densely} networks 
for the classification task directly on the CXRs, fine-tuning their weights pre-trained on ImageNet. We observed that this yields poor classifiers, with less than 50\% accuracy (Items 1-2). 
Concatenating the features from DenseNet and VGG16 and performing heavy hyper-parameter tuning results in a moderately improved performance (Item 3). Overall, the networks were unable to recognize the discriminative regions and performed poorly, due to the small area footprint, long tubular structures that blend into the background, and uneven sample sizes.
 Thus, in further experiments, we treated the ImageNet pre-trained neural networks as feature extractors, feeding the pre-final layer outputs to random forest (RF) classifiers, henceforth abbreviated NN-RF.
Using the original CXR as input to the NN-RF improved accuracy, while the area under ROC (AUC) still remained at 50\% (Items 4-5). 

Next, we processed all the CXRs through the U-Net generating segmentation images. We fed combinations of the segmentation (Seg) and the original CXR image to NN-RFs: Seg alone (Items 6, 8), Seg with CXR as one of the image channels (Items 7, 9) and the original CXR masked to zero out the lowest values of the segmentation output to create a masked CXR image focused on regions of potential CVCs (Items 10-11). These showed considerable improvements in the AUC, while the other metrics (precision, recall, F-score and accuracy) remained primarily unchanged. 

Finally, we extracted the image-processing features describing the size, shape, likelihood based on CVC spatial priors, and relation to chest anatomical elements, as presented in Section~\ref{sec:feature_ip}. The simplest set of features comprising the spatial prior information itself yielded a 12\% increase in AUC (Item 12). The further addition of size and HoG shape features (Item 13) and anatomical relation information (Item 14) improved the classifier performance further, giving 85.2\% accuracy at a precision of 91.6\% and a recall of 89.6\%. 

\subsection{CVC Type Identification}

For the classification problem of identifying the type of CVC present, we again perform a 5-fold cross-validation using a 60-20-20 split for training-validation-testing. We used the 10,746 CXRs which have at least one catheter. 
We report the support-weighted average performance metrics, and the precision and recall for each CVC type in Table~\ref{table:type} using the different methods presented for CVC presence identification (except for DenseNet+VGG concatenation). 

As in the previous classification problem, pre-trained DenseNet and VGG16 are unable to perform well (Item 1-2), though treating these as feature extractors for a RF classification improves performance substantially (Item 3-4). Utilizing the features extracted from fine-tuned VGG performs similarly achieving an average 79.6\% precision, 49.2\% recall and 59.5\% accuracy, while DenseNet performs worse.
Segmentations and CXRs input to the networks for feature extraction marginally improve the average accuracy and AUC (Item 5-10), leaving the other metrics unchanged. 
The image-processing based features perform the best (Item 11-13), reaching an average classification accuracy of 78.2\% at a precision of 95.2\%. 
Our method also achieves lower metric variation as brought out by the standard deviation values of the average accuracy.

\begin{figure*}[t]
\centering
    \includegraphics[width=\textwidth]{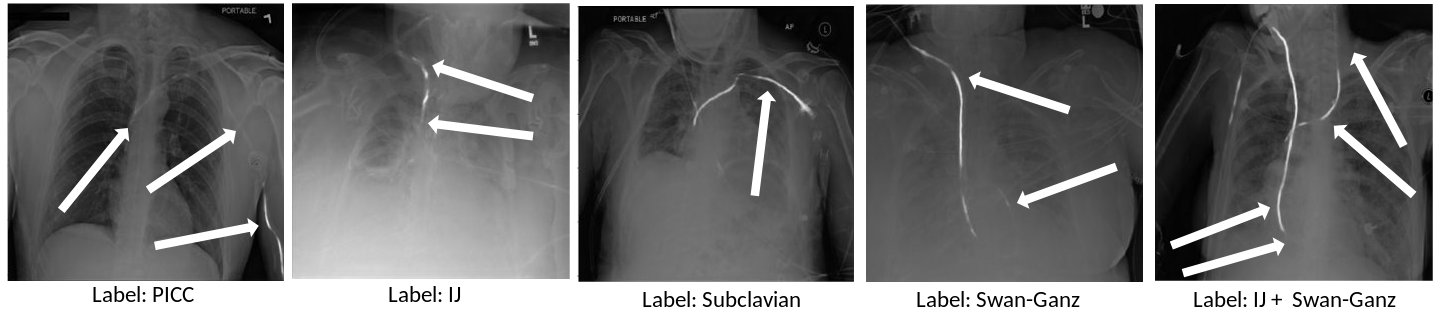}
  \caption{Some examples of CXRs with U-Net output overlayed. Despite the incomplete segmentation by the U-Net, our method correctly labels each of the catheters.}
\label{fig:final_examples}
\end{figure*}

From our best-performing classifier (Item 13), it can be observed that recall for CVCs other than PICC is under 50\%. 
This reveals that the RF classifier performs best on PICC lines, primarily due to the fact that PICC lines make up about 40\% of the CVC images while having the simplest contours. More complex contours like the Swan-Ganz are under-represented and comprise only 2\% of all CVCs, and make the task challenging. This could be overcome by obtaining more data for these classes. 
Some example CXRs with the U-Net output, and the final class prediction are presented in Fig.~\ref{fig:final_examples}. It can be observed that our approach identifies the type of CVC correctly despite the incomplete segmentations from the U-Net. Our method can also handle the presence of multiple CVCs in the same CXR.

\begin{table}[t]
\caption{Results for CVC type identification (Mean $\pm$ standard deviation, P: Precision, R: Recall, AUC: Area under ROC, DN: DenseNet, SP: spatial prior). Best values for each column are in bold. Our algorithm: Rows 11-13.}
\resizebox{\textwidth}{!}{
\begin{tabular}{l|cc|cc|cc|cc||cccc}
 & \multicolumn{2}{c|}{PICC} & \multicolumn{2}{c|}{IJ} & \multicolumn{2}{c|}{SC} & \multicolumn{2}{c|}{SG} & \multicolumn{4}{c}{Weighted Average} \\ \hline
Method & P & R & P & R & P & R & P & R & P & R & Accuracy & AUC \\ \hline
1. DN & 34.6 & 11.6 & 13.2 & 22.2 & 0.0 & 0.0 & 2.6 & 51.4 & 34.2 & 20.4 & 20.0 $\pm$ 0.71 & 51.0 $\pm$ 1.00 \\
2. VGG & 36.6 & 2.8 & 16.6 & 8.8 & 1.2 & 16.8 & 2.8 & 6.2 & 36.0 & 27.6 & 28.0 $\pm$ 0.00 & 48.6 $\pm$ 0.89 \\ \hline
3. DN-RF & 78.4 & 55.0 & 96.6 & 32.4 & \textbf{100.0} & \textbf{33.8} & \textbf{100.0} & \textbf{25.8} & 84.6 & 47.4 & 66.2 $\pm$ 1.10 & 67.0 $\pm$ 0.71 \\
4. VGG-RF & 77.2 & 54.6 & 96.4 & 32.2 & \textbf{100.0} & 33.2 & \textbf{100.0} & \textbf{25.8} & 83.8 & 47.2 & 65.6 $\pm$ 0.89 & 66.8 $\pm$ 0.84 \\ \hline
5. Seg-DN-RF & 77.4 & 64.8 & 84.2 & \textbf{38.2} & \textbf{100.0} & 33.2 & \textbf{100.0} & 24.6 & 80.8 & 55.4 & 68.6 $\pm$ 0.89 & 68.6 $\pm$ 0.89 \\
6. +CXR & 78.2 & 52.2 & 98.2 & 33.2 & \textbf{100.0} & \textbf{33.8} & \textbf{100.0} & 25.2 & 85.0 & 45.8 & 65.4 $\pm$ 1.14 & 66.8 $\pm$ 0.45 \\ \hline
7. Seg-VGG-RF & 76.4 & 63.8 & 82.4 & 37.6 & \textbf{100.0} & 32.6 & \textbf{100.0} & \textbf{25.8} & 79.2 & 54.8 & 67.4 $\pm$ 0.55 & 68.2 $\pm$ 0.84 \\
8. +CXR & 76.2 & 52.6 & 96.2 & 32.8 & \textbf{100.0} & 33.2 & \textbf{100.0} & 23.8 & 83.2 & 45.8 & 64.8 $\pm$ 1.30 & 66.6 $\pm$ 0.55 \\ \hline
9. Mask-DN-RF & 75.6 & 60.0 & 88.6 & 35.6 & \textbf{100.0} & \textbf{33.8} & \textbf{100.0} & 25.2 & 80.6 & 51.6 & 66.6 $\pm$ 0.89 & 67.8 $\pm$ 0.45 \\
10. Mask-VGG-RF & 74.6 & 58.0 & 87.2 & 34.8 & \textbf{100.0} & \textbf{33.8} & \textbf{100.0} & 25.2 & 79.2 & 50.2 & 65.4 $\pm$ 1.14 & 67.2 $\pm$ 0.84 \\ \hline
11. Seg-SP-RF  & 88.6 & 75.4 & 95.8 & 33.4 & \textbf{100.0} & 31.6 & \textbf{100.0} & \textbf{25.8} & 91.2 & 61.6 & 75.6  $\pm$ 0.55 & 70.0  $\pm$ 0.71 \\
12. + HoG  & 93.0 & 78.8 & 99.0 & 32.4 & \textbf{100.0} & 33.2 & \textbf{100.0} & \textbf{25.8} & 95.2 & \textbf{63.8} & 78.2 $\pm$ 0.84 & 70.8 $\pm$ 0.84 \\
13. + Anatomy  & \textbf{93.0} & \textbf{78.8} & \textbf{99.6} & 32.2 & \textbf{100.0} & \textbf{33.8} & \textbf{100.0} & \textbf{25.8} & \textbf{95.6} & 63.6 & \textbf{78.4} $\pm$ \textbf{0.55} & \textbf{71.0} $\pm$ \textbf{0.71} \\ \hline
\end{tabular}}
\label{table:type}
\end{table}

\section{Conclusions}
In this paper, we have addressed, for the first time, the detection and classification of central venous catheters in chest X-rays. Due to the small footprint of these devices, it is difficult for deep learning networks to detect these structures based on whole image input.
Our method presents a robust solution to this problem by using deep learning for an approximate segmentation of these structures followed by conventional machine learning on features from these regions incorporating spatial priors from distribution atlases extracted from CXR images. Techniques to better incorporate handcrafted features, including anatomical features and spatial priors, into neural networks will be explored in the future. 

\bibliographystyle{splncs18}

\end{document}